\documentstyle{article}[12pt]
\def\ao{\hat{a}}
\def\co{\hat{a}^\dagger}
\def\aot{\hat{a}^{2}}
\def\cot{\hat{a}^{\dagger 2}}

\begin{document}
%\draft
\title{Nonlinear Jaynes-Cummings model of  atom-field interaction}
\author{S. Sivakumar
\footnote{Email: siva@igcar.ernet.in} \\
Indira Gandhi Centre for
Atomic Research, Kalpakkam 603 102 India\\
}
\maketitle
\\
{\bf Short title}\\
\begin{center}
\title{Nonlinear Jaynes-Cummings model of  atom-field interaction}
\end{center}
\begin{abstract}
   Interaction of a two-level atom with a single mode of electromagnetic
 field including Kerr nonlinearity for the field and intensity-dependent
 atom-field coupling is discussed.  The Hamiltonian for the atom-field
 system  is written in terms of the elements of a closed algebra, which has
 SU(1,1) and Heisenberg-Weyl algebras as limiting cases. Eigenstates and
 eigenvalues of the Hamiltonian are constructed.
 With the field being in a  coherent state initially, the dynamical behaviour
 of atomic-inversion, field-statistics and uncertainties in the field
 quadratures are studied. The appearance of nonclassical features during the
 evolution of the field is shown.   Further, we explore the overlap of
 initial and time-evolved field states.
\end{abstract}
\newpage
\section{Introduction}

         Interaction of a single mode of electromagnetic field with a
two-level atom is the simplest problem in matter-radiation coupling.
A model for the interaction, introduced by Jaynes and Cummings \cite{jaynes},
treats the atom as a dipole placed in an external field.
The Jaynes-Cummings (JC) model has provided a lot of impetus for
theoretical explorations and experimental verifications
\cite{rev1,rev2,rev3,rev4,rev5}.
The Hamiltonian for the model is
\begin{equation}
H_{JC} = H_0 +g(\co\sigma_- + \ao \sigma_+ ).
\end{equation}
Here $H_0$, the Hamiltonian for the atom and field in the absence
of interaction, is $\omega\co\ao+{1\over 2}\nu\sigma_z$.  The
atomic transition frequency is $\nu$, the field frequency is
$\omega$ and $g$ is the coupling constant. We denote the creation
operator by $\co$, annihilation operator by $\ao$, and their
action on the basis states of the harmonic oscillator are\\
\begin{eqnarray}
\ao\vert n\rangle&=&{\sqrt n}\vert n-1\rangle,\\
\ao\vert 0\rangle&=&0,\\
\co\vert n\rangle&=&{\sqrt{n+1}}\vert n+1\rangle.
\end{eqnarray}
The atom has two-levels, $\vert g\rangle$ and $\vert e\rangle$, the ground
and excited states.  The operators $\sigma_z$, $\sigma_+$ and $\sigma_-$ act
on the states as given below:
\begin{eqnarray}
\sigma_z \vert g\rangle&=&- \vert g\rangle \\
\sigma_z \vert e\rangle&=&\vert e\rangle \\
\sigma_\pm \vert g\rangle&=&{1\pm 1\over 2}\vert e\rangle \\
\sigma_\pm \vert e\rangle&=&{1\mp 1\over 2}\vert g\rangle \\
\end{eqnarray}

Atomic inversion, defined as the   difference in probabilities for
the atom to be in the excited and ground states, as predicted by JC model
is a sum of quasiperiodic functions with incommensurate frequencies.
The model predicts collapses, revivals and ringing revivals in the
time-development of atomic inversion\cite{cummings,
stenholm, meystre,eberly, mvs}
The revival phenomenon is entirely quantal,  and hence the model is
very important for testing predictions of quantum theory.
The model has been generalized  in very many ways. We list the respective
Hamiltonian for some of the models:\\
1) Buck-Sukumar model\cite{buck}
\begin{equation}
H_{IC}=H_0 + g({\sqrt{\co\ao}} \ao \sigma_+ + \co{\sqrt{\co\ao}}\sigma_-)
\end{equation}
With this particular form of intensity-dependent coupling the
atomic-inversion is a sum of periodic functions.  This model is very
interesting as it can be written as a combination of generators of SU(1,1)
algebra\cite{buzek}.  Generalization to two-photon case, where two-photons
are absorbed or emitted in atom-field interactions, has been done.  Once
again the SU(1,1) algebraic structure in the model is used to solve the
problem \cite{gerry}. \\ \\
2) Kerr-nonlinearity\cite{wernerrisken,gora,puri}\\
\begin{equation}
H_{Kerr}=H_{JC}+ \chi\cot\aot
\end{equation}
This is an effective Hamiltonian for a system in which the
electromagnetic field mode is excited in a Kerr medium.  The
medium is modelled as an
anharmonic oscillator\cite{yurke, agarwal}.\\ \\
3)Dicke-Tavis-Cummings model\cite{dicke,tavis}\\ \\
 In this model, the interaction between field and a group of two-level
 atoms is considered and the Hamiltonian is
\begin{equation}
H_{DTC}=\omega\co\ao+{\nu\over 2}\sum_i\sigma_{i,z}+\hbox{Interaction part}
\end{equation}
4) Nonlinear Jaynes-Cummings model\\ \\
By including the motion of atom in the external field, the coupling is made
position dependent.  This offers enormous possibilities to tailor the
form of atom-field interaction \cite{vogel,liu}.  The general form for
the Hamiltonian is
\begin{equation}
H_{NL}=H_0+g(f(\co\ao){\ao}^m+\hbox{adjoint})
\end{equation}
Here $f(\co\ao)$ is an operator-valued function of the
number operator $\co\ao$ and $m$ is an integer.

        The objective of the present paper is to study the dynamics of a
  two-level atom interacting with a single mode of electromagnetic field.
 The interaction is governed by the Hamiltonian
\begin{equation}\label{hamil}
H = \omega\left[\co\ao+{r\over 2}\sigma_z\right] + \chi\cot\aot
+g\omega( {\sqrt{1+k\co\ao}} \ao \sigma_+ + \co {\sqrt{1+k\co\ao}}
\sigma_-).
\end{equation}
This Hamiltonian is an example for nonlinear JC model including Kerr term.
Note that we have scaled the coupling constant  $g$ by $\omega$.
The parameter $r$ is $\nu\over\omega$ and the coupling constant for
the Kerr term is $\chi=k\omega$.
This will simplify the expressions we derive in sequel.  In particular,
we set $\omega=1$ which amounts to studying a new Hamiltonian $H\over\omega$.
The speciality of the Hamiltonian $H$ is that it becomes $H_{JC}$ when the
parameter $k$ is set equal to zero.  Further, the usual
Holstein-Primikoff realization is obtained when $k = 1$ so that
the interaction is approximately given by
$g({\sqrt{\co\ao}}\ao\sigma_+ + \hbox{adjoint})$,
which is same as the interaction studied in Ref.\cite{buzek}.  Yet another
form of interaction occurs when $k\ll 1$. If the photon number distribution
is such that $k n\ll 1$ for all $n$ under the peak of the distribution, then
 Eq. \ref{hamil} leads to
\begin{equation}
H\rightarrow \omega\left[\co\ao+k\cot\aot+g[(1+\co\ao/2)\ao\sigma_+ +
\hbox{adjoint}]\right].
\end{equation}
Further, when $k\ll 1$ so that we can neglect $k n$ in comparison to
unity but retain  $k n^2$,  we arrive at $H_{JC}$ with an additional
Kerr term.  This system has been studied in Ref.\cite{jex}.
Many well studied systems are thus special cases of the Hamiltonian
considered for discussion in the present paper.
The organization of the paper is as
follows.  In section II we study the algebraic aspects of the Hamiltonian
and construct the eigenvectors and eigenvalues.  Section III is devoted
to study the atomic inversion and approximate expressions are obtained for
first collapse and revival periods when the field is in a coherent state.
The time-development of field statistics is discussed in section IV and the
results are summarized in Section V.

\section{Model Hamiltonian and its properties}

        In this section we study the algebraic aspects of the generalized
 Hamiltonian given in Eq. (\ref{hamil}).  We introduce a set of 
 operators which are closed under commutation. This  algebraic
 structure is exploited to  determine the time-evolution operator to evolve
 the initial state of the atom-field system.  The eigenvalues and the
 corresponding eigenvectors are constructed.

\subsection{Eigenvalues and eigenvectors}

        The Hamiltonian given in Eq.\ref{hamil} is written as
\begin{equation}
H=\omega K_+ K_- + {\nu\over 2} \sigma_z + g(K_+ \sigma_- + K_- \sigma_+),
\end{equation}
wherein we have set $K_-={\sqrt{1+k\co\ao}}\ao$ and
$K_+=\co{\sqrt{1+k\co\ao}}$.
  Further, we assume that $k$ is nonnegative and restricted to take values
  less than or equal to unity. Formally, the Hamiltonian $H$ has the same
  structure as $H_{JC}$  with $\ao$ and $\co$ replaced by $K_-$ and $K_+$.
  However, the former corresponds to Kerr-type medium with
 intensity-dependent coupling for atom-field interaction.   The difference is
 very clear in the commutation relations among the operators.  From the
 realisation of $K_-$ and $K_+$ in terms of $\co$ and $\ao$, we arrive at
\begin{eqnarray}
[ K_- , K_+ ] &=& 2 K_0,\cr
[ K_0 , K_{\pm} ] &=& \pm k K_\pm .
\end{eqnarray}
The operator $K_0$ is $k\co\ao+{1\over 2}$.  Thus, the operators $K_-,K_+$ and
$K_0$ form a closed algebra.  It worth noting that the commutation relations
define the SU(1,1) algebra when $k=1$.  On the other hand, 
to  get the well-known
 Heisenberg-Weyl algebra generated by $\co,\ao$ and the identity $I$ we set
 $k=0$.   Two
different algebras are realized depending on the value of $k$
and hence the algebra of $K_-,K_+$ and $K_0$ is said to be an
``interpolating
 algebra". An  invariant operator, which
commutes with $K_\pm$ and $K_0$,
 for the algebra is given  by $K_0^2-(k/2)(K_-K_++K_+K_-)$.
The coherent states
corresponding to this algebra and their Hilbert space properties are known
\cite{siva}.

        The atom-field evolution is studied in the space of
 $\vert e,n\rangle$ and  $\vert g,n\rangle$, where $n=0,1,2,\cdot\cdot\cdot$.
 The state $\vert e,n\rangle$ means that the atom is in the excited state
 $\vert e\rangle$ and the field in
 the $n^{\hbox{th}}$ excited state $\vert n\rangle$.  The states
 $\vert e,n\rangle$ and $\vert g,n\rangle$ are eigenstates of
 $\omega K_+K_- + {\nu\over2}\sigma_z$ and the respective eigenvalues are
 ${\mathcal E}_{e,n}=(n+k n^2-k n)\omega+{\nu\over 2}$ and
 ${\mathcal E}_{g,n}=(n+k n^2-k n)\omega-{\nu \over 2}$.
 The Hamiltonian $H$ admits a constant of motion N such that the commutator
 $[N,H]$ vanishes.  Explicitly,  $N=K_+K_-+2K_0\sigma_+\sigma_-$.  Note that
 when $k=0$, $N$ becomes $\co\ao+\sigma_+\sigma_-$, the constant of motion
 for $H_{JC}$.

        The interaction part of $H$ is such that the state $\vert e,n\rangle$
is taken to $\vert g,n+1\rangle$ and {\it vice versa}, during the evolution
of the atom-field system.  Thus, the entire Hilbert space is split into
subspaces spanned by $\vert e,n\rangle$ and $\vert g,n+1\rangle$ and
the dynamics confined to individual subspaces. In one such subspace,
specified by the value of  $n$, the Hamiltonian matrix is
\begin{equation}
H=\pmatrix { (n+{1\over 2}+k n^2 -k n)\omega +\Delta/2 &
g\sqrt{(1+k n)(1+n)} \cr
g\sqrt{(1+k n)(1+n)} &
(n+{1\over 2}+k n^2 +k n)\omega - \Delta/2\cr}.
\end{equation}
The detuning parameter $\Delta$ is $(r-1)\omega$.  The eigenvalues of the
Hamiltonian are
\begin{equation}
E_{\pm,n}=(k n^2+ n+{1\over 2})\omega
\pm{1\over 2}\sqrt{(\Delta-2 k \omega n)^2+4 g^2\omega^2 (1+k n)(1+n)}.
\end{equation}
and the  corresponding eigenvectors are
\begin{eqnarray}
\vert+,n\rangle=\cos\theta_n\vert e,n\rangle+\sin\theta_n\vert g,n+1\rangle,\\
\vert-,n\rangle=\sin\theta_n\vert e,n\rangle-\cos\theta_n\vert g,n+1\rangle.
\end{eqnarray}
The expansion coefficients are
\begin{eqnarray}
\cos\theta_n={2g\omega\sqrt{(1+n)(1+kn)}\over
\sqrt{(\Omega_n-\Delta_n)^2+4g^2\omega^2(1+n)(1+kn)}}\\
\sin\theta_n={\Omega_n-\Delta_n\over
\sqrt{(\Omega_n-\Delta_n)^2+4g^2\omega^2(1+n)(1+kn)}},
\end{eqnarray}
in which we have set $\Delta_n=\Delta-2kn\omega$ and
$\Omega_n=\sqrt{\Delta_n^{ 2}+4g^2\omega^2(1+n)(1+kn)}$.

          The energy difference  between the levels $E_{+,n}$ and $E_{-,n}$
  is  $\sqrt{\Delta_n +4g^2\omega^2(1+kn)(1+n)}$.
The minimum of the separation occurs when $\Delta$ equals
$2kn\omega$ and the corresponding difference is
$2g\omega\sqrt{(1+kn)(1+n)}$.
  In Fig. 1  we have plotted the
energy eigenvalues $E_+$ and $E_-$ as a function of $\Delta$.  The
dashed lines represent the eigenvalues when $g=0$, {\it i.e.},
${\mathcal E}_{e(g),n}$. In this case the eigenvalues cross each
other as $\Delta$ increases from negative to positive  values. The
continuous lines represent the energy eigenvalues for $g=10^{-3}$.
 The diverging eigenvalue separation beyond the minimum
separation indicates ``level repulsion'' in the eigenvalues of the
dressed atom.  The effect of nonzero $k$ is to shift the value of
$\Delta$ at which the minimum separation or the crossing occurs.
If $k=0$, the minimum as well as the crossing occur at $\Delta=0$.

\subsection{Evolution of atom-field state}

        To understand the dynamics of the atom-field system, we
solve for the state of  the system in interaction picture,
where the evolution equation is
\begin{equation}\label{IPH}
i{\partial\vert\psi\rangle\over\partial t}=\tilde{V}\vert\psi\rangle.
\end{equation}
Here $\tilde{V}$ is the transformed interaction given by
\begin{equation}
\tilde{V}= g\exp[it(\omega K_+K_- +{\nu\over 2}\sigma_z)]
(\sigma_- K_++\sigma_+ K_-)
\exp[-it(\omega K_+K_- +{\nu\over 2}\sigma_z)].
\end{equation}
The effect of the transformation on the interaction term is obtained from
the following results:
\begin{eqnarray}
\exp(it\omega K_+K_-)K_+\exp(-it\omega K_+K_-)&=&K_+\exp(it\omega K_0),\\
\exp({it\nu\over 2}\sigma_z)\sigma_-\exp(-{it\nu\over 2}\sigma_z)&=&
\exp(-it\nu)\sigma_-,
\end{eqnarray}
and their adjoints.  Note that the ordering  of operators should
be maintained in the rhs of the first of the results.  Using these relations,
the interaction picture Hamiltonian is written as
\begin{equation}
\tilde{V}=g(K_+\sigma_-\exp[it(2\omega K_0-\nu)]+\hbox{adjoint}.
\end{equation}
At any time $t$, let the state of the atom-field system  be represented as
\begin{equation}
\vert\psi(t)\rangle=\sum_{n=0}^\infty C_{e,n}(t)\vert e,n\rangle +
C_{g,n}(t)\vert g,n\rangle.
\end{equation}
The coefficients $C_{e,n}(t)$ and $C_{g,n}(t)$, determined in
terms of their initial values by the evolution equation
Eq.\ref{IPH}, are
\begin{eqnarray}\label{tdsf}
\exp\left({-i\Delta_n t\over 2}\right) C_{e,n}(t)&=&
\left[ \cos({\Omega_n t\over 2})-{i\Delta_n\over\Omega_n}
\sin\left({\Omega_n t\over 2}\right)\right] C_{e,n}(0)-\nonumber\\& &
{2ig\sqrt{(n+1)(1+kn)}\over\Omega_n}
\sin\left({\Omega_n t\over 2}\right)C_{g,n+1}(0),\\
\exp\left({i\Delta_n t\over 2}\right) C_{g,n+1}(t)&=&
\left[ \cos({\Omega_n t\over 2})+{i\Delta_n\over\Omega_n}
\sin\left({\Omega_n t\over 2}\right)\right] C_{g,n+1}(0)\nonumber\\& & -
{2ig\sqrt{(n+1)(1+kn)}\over\Omega_n}
\sin\left({\Omega_n t\over 2}\right)C_{e,n}(0).
\end{eqnarray}
The  Rabi frequency $\Omega_n$ is
$\sqrt{\Delta_n^{2}+4g^2\omega^2(1+kn)(1+n)}$.
The dependence of $\Omega_n$  on $n$ such is that there is a minimum value
for the Rabi frequency when $n$ satisfies $\Delta=2kn\omega+
g^2\omega^2(1+k+2kn)( k\omega)^{-1}$, provided $k\ne 0$.   The
variation of $\Omega_n$ with respect to $n$ is shown in Fig. 2.   In the case
of $H_{JC}$,   the Rabi frequency varies linearly with $n$ and hence there
is no minimum.  The existence of a minimum Rabi frequency has important
consequences for the dynamics of atomic-inversion, squeezing,
 photon-statistics, {\it etc.} and they are discussed in the following
  sections.\\

Let the Rabi frequency attain its
minimum for some specific value of $n$, denoted by $\bar{n}$.  Therefore,
we have
\begin{equation}
\Delta=\Delta_c=2k\omega\bar{n}+{g^2(1+k+{\bar n} )\over k\omega}.
\end{equation}
The $n$ dependence of Rabi frequency, for values of $n$ close to $\bar{n}$,
is obtained by Taylor expanding $\Omega_n$  around $\bar{n}$, upto second
 order. The resultant expression is
\begin{equation}\label{appomega}
\Omega_n=\Omega_{\bar{n}}+
{ 2 {(n-\bar{n})}^2 (k^2\omega^2+kg^2)\over\Omega_{\bar{n}}}
\end{equation}
In Fig. 2 the values predicted by the  approximate expression for $\Omega_n$
are compared with those of the exact expression. It is clear that
for the chosen values of $g$, $k$ and $\bar{n}$, the values match very well
for those values of $n$ under the peak of the photon number distribution.
  Quantitatively, the fractional difference is less than 3\%.

\section{Evolution of atomic inversion}

        In the previous section we constructed the complete state of the
atom-field system in the dressed atom basis in interaction
picture.  The state $\vert\phi(t)\rangle$ in the Schr{\"o}dinger
picture is easily obtained by premultiplying the interaction
picture state function $\vert\psi(t)\rangle$ by $\exp(i(K_-K_+
+\nu/ 2)t)$.  In this section, we study the temporal evolution of
atomic inversion.\\

        The time-dependent state vector $\vert\psi(t)\rangle$ of the system
is determined completely once the coefficients $C_{e,n}(t)$ and $C_{g,n}(t)$
are known, which, in turn, are specified by their initial values.  For
instance, if the atom is initially in the excited state, we have
$C_{g,n}(0)=0$ and $C_{e,n}(0)=\langle n\vert\psi(0)\rangle$.  The probability
 that the atom is in the excited state, irrespective of the state of the field,
 is $\sum_{n=0}^\infty P_n \vert C_{e,n}(t)\vert^2$ and
 to be in the ground state is $\sum_{n=0}^\infty P_n\vert C_{g,n}(t)\vert^2$.
The photon number distribution of the field is $P_n$. The
difference of these two probabilities is atomic inversion.  For
the specified initial condition, namely, the atom is initially in
the excited state, the atomic inversion $W(t)$  is
\begin{equation}
W(t)=1+\sum_{n=0}^\infty P_n\left[
{4g^2\omega^2(1+n)(1+kn)\over\Omega_n^2}(\cos(\Omega_nt)-1)\right],
\end{equation}
and time-dependent part of  $W(t)$ is
\begin{equation}\label{wtime}
W_T(t)=4g^2\omega^2\sum_{n=0}^\infty
P_n {(1+n)(1+kn)\over\Omega_n^2}\cos(\Omega_n t).
\end{equation}

        The quantity $W_T$ exhibits rich structure in its evolution.  It
exhibits collapse  and revivals when the initial photon
distribution is taken to be a Poissonian distribution of mean
photon number $\bar{n}$, which corresponds to the field being in a
coherent state $\vert\sqrt{\bar{n}}\rangle$. The photon number
distribution for the state is
$P_n=\exp(-\bar{n}){{\bar{n}}^n\over{n!}}$. This distribution
has a single peak and the standard deviation is $\bar{n}$. Hence,
the major contribution to the sum in Eq.\ref{wtime} comes from a
few terms with $n$ around the peak. With this choice of $P_n$, we
have plotted in  Fig. 3a the evolution of $W_T$ as a function of
time.  The time required for the first collapse  and the following
revival, denoted by $T_C$ and $T_R$ respectively,  can be
estimated approximately.  For the revival to occur, the terms
corresponding to those $n$ around the peak, should be in phase.
Thus, we require $T_R (\Omega_{\bar{n}+1}-\Omega_{\bar{n}})$ be
equal to $2\pi$.  The difference of the nearest-neighbour Rabi
frequencies is

\begin{equation}
\Omega_{\bar{n}+1}-\Omega_{\bar{n}}={2A\bar{n}+A+B\over{2\Omega_{\bar n}}}.
\end{equation}
The constants $A$ and $B$ are $4(k^2\omega^2+kg^2)$ and
$4(g^2+kg^2-\Delta k\omega)$ respectively.  The revival time $T_R$
is $4\pi\Omega_{\bar n}\over{2A\bar{n}+A+B}$.\\

        For the inversion to collapse, the terms in the sum on the rhs of
Eq.\ref{wtime} should be uncorrelated.  Since the width of a Possionian
distribution is where the probability $P_n$ is appreciable, the condition
for collapse is written as
$T_C ({\Omega_{\bar{n}+\sqrt{\bar{n}}}}-
{\Omega_{\bar{n}-\sqrt{\bar{n}}}})=1$.  If $\bar{n}$ is large, the expression
for $T_C$ is $T_R\over 4\pi\sqrt{\bar{n}}$.

The function $W_T$ exhibits rich features when detuning is close to $\Delta_c$.
In Fig. 3.  we have shown the behaviour of $W_T$ for three different values of
$\Delta$.  The values for  the parameters are $g=10^{-3}$, $k=10^{-4}$,
$\omega=1$, $\bar{n}=30$ and the corresponding $\Delta_c$ is 0.01606.
The values are so chosen that the Rabi frequency
attains its minimum when $n$ is near $\bar n$, the average photon number.
The evolution of $W_T$ with $\Delta=.01 <\Delta_c$ is shown as the top most
figure, marked $(a)$ in Fig. 3.  The figure $(b)$ corresponds to
$\Delta=\Delta_c$
and the one marked $(c)$ is for $\Delta=.22 > \Delta_c$.  The
envelope of
$W_T$ when detuning equals  $\Delta_c$, is distinct with
structures repeating without much distortion.  This should be compared with
the top and bottom figures, which correspond to $\Delta\ne\Delta_c$,
 which exhibit random oscillations and do not have neat envelope.

        As noted in Section II, SU(1,1) algebra is realized in terms
of $K_\pm$ and $K_0$ when $k=1$.  If we consider resonant
interaction $(r=1)$, then $W_T$ can be estimated approximately for
the SU(1,1) case. The field is taken to be in a coherent state
$\vert\alpha\rangle$ such that $\vert\alpha\vert = {\sqrt {\bar
n}} \gg 1$.  We use  $(1+n)(1+kn)\approx n^2$ and $g^2\ll 1$ to
arrive at $\Omega_n=2 n\omega$.  With these approximations,
 we arrive at
\begin{equation}
W_T=g^2\exp(\bar{n}(\cos(\omega t)-1))\cos(\bar{n}\sin(\omega t)),
\end{equation}
in which we have used $(1+n)(1+kn)\approx{\Omega_n}^2$.
The magnitude of $W_T$ is negligible in this case and so there is no
 perceptible collapse or revival.  This is due to the presence of
 Kerr term and the fact that it dominates
 over $\co\ao$ in the Hamiltonian.  However,  we point  out that
 collapses and revivals are present in $W_T$ if $\chi\ll\omega$.

\section{Dynamics of field properties}

        In the previous section,  we studied the dynamics of the two-level
atom, in particular, the atomic inversion.
In the present section, we explore the temporal
behaviour of field statistics and field amplitudes.  As in the previous
section, the field is initially in a coherent state of complex amplitude
$\alpha$ and the atom is taken to be in its excited state.  With these
initial conditions, the probability distribution of photons at time $t$ is
\begin{eqnarray}
P(n,t)&=&\vert C_{e,n}(t)\vert ^2 + \vert C_{g,n}(t)\vert ^2\nonumber\\
&=& {P_n\over 2}
\left[
1+{\Delta_n^{2}\over\Omega_n^2}
 + {4 g^2\omega^2 (1+n)(1+kn)\over \Omega_n^2}\cos(\Omega_n t)
\right]
\end{eqnarray}
Coherent states are the wavepackets whose behaviour is closest to
that of a classical particle and hence are called classical
states.  Nevertheless, during their evolution in time the states
may not be classical.  The photon statistics of coherent states is
Poissonian. Any deviation from this behaviour is characterized by
Mandel's Q parameter defined as
$\langle\co\ao\co\ao\rangle-\langle\co\ao\rangle^2\over\langle\co\ao\rangle$.
Using the time-dependent probability distribution $P(n,t)$, the
expectation values in the expression for Q parameter are
\begin{equation}
Q={\sum_{n=0}^\infty n^2 P(n,t)-{[\sum_{n=0}^\infty P(n,t)n}]^2\over
\sum_{n=0}^\infty P(n,t) n}
\end{equation}
 For coherent states, Q is unity.  Any value of Q less than unity is
 nonclassical.  In Fig. 4  the time evolution of Q parameter
  for an initially coherent
 state and $\Delta=0.01<\Delta_c$ is given. The emergence of nonclassical
 behaviour ($Q < 1$) is seen. Though not shown in figure, we point out that
 for  $\Delta=\Delta_c=0.016061$, the statistics does not become
 sub-Poissonian. When $k=1$ and $\Delta=0$, the time-dependent part of
 $P(n,t)$ is of negligible magnitude and  so $Q$ does not evolve in time.

\subsection{Squeezing}
        We define the field amplitudes to be
\begin{eqnarray}
X(t)= {\co(t)+\ao(t)\over\sqrt{2}},\\
Y(t)=i{\co(t)-\ao(t)\over\sqrt{2}}.\\
\end{eqnarray}
These amplitudes satisfy the commutation relation $[X,Y]=i$ and hence
they satisfy $(\delta X)(\delta Y)=1/2$.  The symbol $(\delta X)$ stands
for the expression $\sqrt{\langle X^2\rangle-\langle X\rangle^2}$,
the variance in $X$ for a given field state.
For coherent states of any amplitude $\alpha$,
variances in $X$  and $Y$ are the same
and equal to 1/2.  A state is nonclassical if $(\delta X)$ is less than 1/2,
the coherent state value.  Using the time-dependent state function given in
 Eq. (\ref{tdsf}), the variances in the field amplitudes are given by
\begin{equation}
(\delta X)^2={1\over 2}\left[1+\langle 2\co\ao+\cot+\aot\rangle-
\langle\ao\rangle^2-\langle\co\rangle^2-2\langle\co\rangle\langle\ao\rangle
\right],
\end{equation}
and
\begin{equation}
(\delta Y)^2={1\over 2}\left[1+\langle 2\co\ao-\cot-\aot\rangle+
\langle\ao\rangle^2+\langle\co\rangle^2-2\langle\co\rangle\langle\ao\rangle
\right].
\end{equation}
The expectation values of various operators in these expressions are
\begin{eqnarray}
\langle\ao\rangle&=&\sum_{n=0}^\infty\sqrt{n+1}
\left[ C^*_{e,n}C_{e,n+1}+C^*_{g,n}C_{g,n+1}\right],\\
\langle\aot\rangle&=&\sum_{n=0}^\infty\sqrt{(n+1)(1+kn)}
\left[ C^*_{e,n}C_{e,n+2}+C^*_{g,n}C_{g,n+2}\right],\\
%end{eqnarray}
\hbox{and}\nonumber\\
%\begin{equation}
\langle\co\ao\rangle &=& \sum_{n=0}^\infty n
\left[ C^*_{e,n}C_{e,n}+C^*_{g,n}C_{g,n}\right].
%\end{equation}
\end{eqnarray}
The expectation values $\langle\co\rangle$ and $\langle\cot\rangle$ are the
complex conjugates of $\langle\ao\rangle$ and $\langle\aot\rangle$
respectively.

        The evolution of  $\delta X$ is shown in Fig. 5.  As the
field evolves in time, the variance in $X$ falls below 0.5
indicating that the quadrature exhibits squeezing.  As a
consequence of uncertainty relation, the   $Y$ quadrature does not
show squeezing.  However, when we set $\alpha=i{\sqrt 30}$, the
situation is reversed. In this case, squeezing is possible in $Y$
and not in $X$. An interesting feature is that the uncertainty in
$Y$ with $\alpha=i\sqrt{30}$ is same as in $X$ when
$\alpha=\sqrt{30}$.  The reason for this is that the field states
are of the form $\sum_{n=0}^\infty B_n\alpha^n\vert n\rangle$
wherein the coefficients $B_n$ are real for all $n$.  For such
superpositions, the uncertainty in $X$ for a given $\alpha$ is
same as that in $Y$ with $\alpha$ replaced by
$i\alpha$\cite{siva}.

\subsection{Overlap of initial and time-evolved states}

        A quantity of interest is the overlap of the state of the
atom-field at time $t$ and that at $t=0$.  With the same initial conditions
for the atom-field state as in the previous section, the overlap is
\begin{equation}
\vert\langle\psi(0)\vert\psi(t)\rangle\vert^2=\exp(-\vert\alpha\vert^2)
\left\vert\sum_{n=0}^\infty{\vert\alpha\vert^{2n}\over n!}
\left[ \cos({\Omega_n t\over 2}) + i{\Delta_n\over\Omega_n}
\sin({\Omega_n t\over 2})\right]
\right\vert^2.
\end{equation}
The numerical value of the overlap lies between zero and unity.  It is seen
from Fig. 6 that the overlap becomes zero at longer times.
In other words, the time-evolved state is almost orthogonal to initial state.
For short durations, an approximate expression for the decay of overlap is
derived by replacing
 expression $\exp(-\bar{n}){\bar{n}\over n!}$ with the Gaussian distribution
$\exp(-{(n-\bar{n})^2\over 2\bar{n}})$ and  the sum over $n$ by
integration. Further, we set $\Omega_n=\Omega_N+(N-n)
{\Omega^{\prime}_N}+
{\Omega^{\prime\prime}_N}(N-n)^2$, with the
assumption that $N$ is close to but not less than $\bar{n}$.
Here, prime denotes taking
derivative with respect to $n$ and the suffix represents the value  of
$n$ where the derivative is evaluated.
With these
approximations we get, after neglecting oscillatory terms,
\begin{eqnarray}\label{appolap}
\sum_{n=0}^\infty\exp(-{(n-\bar{n})^2\over 2\bar{n}})
 \cos({\Omega_n t\over 2}) &=&
 \hbox{Re}\int_{0}^\infty
\exp(-{(n-\bar{n})^2\over 2\bar{n}}+i{\Omega_n t\over 2})\hbox{d}n
\nonumber\\
&\propto&  \left[
1+N^2{\partial^2\Omega_n\over\partial^2 n}\vert_{n=N}{t^2\over 4}
\right] ^{-{1\over 4}}\nonumber\\
& &\times
\exp\left[ -
N^2{\partial\Omega_n\over\partial n}\vert_{n=N}{t^2\over 4}
\right]
\end{eqnarray}
Similar expression can be derived for summation with $\sin({\Omega_n t\over
 2})$.
The above expression indeed predicts that the overlap function decays with
time.   When $N=\bar{n}$, the
first derivative of $\Omega_n$ vanishes and  the exponential
term in the envelope is absent. Consequently the decay is slower.
However, if
the photon number distribution of the field is very broad, it is incorrect
to truncate the Taylor series and the expression in Eq. (\ref{appolap})   is
invalid.

\section{Summary}

        The Hamiltonian that contains the usual and intensity-dependent
 (including Kerr term) JC models
 as limiting cases has been constructed.  The nondissipative
dynamics dictated by the generalized Hamiltonian $H$ is completely
solvable. The eigenvalues of the Hamiltonian, with Kerr term and
intensity-dependent coupling, exhibits level-repulsion.  In the
case of nonvanishing Kerr term, the  Rabi frequency, considered as
a function of the photon number $n$, attains a minimum. The
dynamical behaviour of atomic-inversion, when detuning is so
chosen to have the minimum Rabi frequency, exhibits
superstructures, which is absent in the usual JC model.  The
overlap of the initial coherent state and  the time-evolved state
decays with time. In the language of inner product, the  initial
coherent state is almost orthogonal to the time-evolved state. The
results, except the occurrence of superstructures in atomic
inversion, a consequence of nonvanishing Kerr term, go over to
those of the JC model, in the limit of $k\rightarrow 0$.\\

         We note that the formal equivalence between $H_{JC}$ and $H$ is
 obtained by identifying  $K_+$ with $\co$ and $K_-$ with $\ao$.  This,
 in conjunction with the fact both the sets of operators $\{\co,\ao,I\}$
 and  $\{K_\pm,K_0\}$ are closed under commutation,
 implies that the expansion coefficients $C_{{e/g},n}(t)$
 for the evolution governed by $H$ are obtained from the
corresponding expressions for the usual JC model
by replacing $\Delta$ by $\Delta_n$ and $g$ by $g\sqrt{1+kn}$.  
Hence, all those physical quantities, like the atomic inversion, quadrature
fluctuations, {\it etc.},  computed in terms of  the expansion
coefficients are derivable from those of the JC model.

\newpage
\begin{center}
{\bf List of Figures}
\end{center}
Fig. 1  Dependence of eigenvalues $E_+$ and $E_-$ on detuning
$\Delta$. The continuous curve corresponds to $g=0.1$ and the
dashed curve corresponds to $g=0$. Here $k=.1$ and $\bar{n}=30$.
The dashed curves with positive and negative slopes correspond
respectively to ${\mathcal E}_{e,n}$ and  ${\mathcal E}_{g,n}$.
Lower part of the figure is for $n=1$ and upper part for $n=2$.
\\ \\
Fig. 2 Variation of $\Omega_n$ with $n$. We have set $g =10^{-3}$,
$k=10^{-4}$, $\bar{n}=30$ and $\Delta=0.016061$. The approximate
and actual Rabi frequencies are compared in the upper figure.
Dotted line corresponds to the approximate expression in
Eq.\ref{appomega} and continuous curve corresponds to exact
expression.  Values of $k$, $g$, $\bar{n}$ and $\Delta$ are
$10^{-4}$, $10^{-3}$, $30$ and $0.016061$ respectively. The bottom
curve shows the photon number distribution for the coherent state
$\vert\alpha={\sqrt 30}\rangle$.
 \\ \\
Fig. 3 Time-dependent part of atomic inversion.  Case (a) corresponds to
$\Delta=0.01 < \Delta_c$. Case (b) is for $\Delta=\Delta_c=0.016061$ and
Case (c) refers to $\Delta=0.022>\Delta_c$.  Here Case (b) shows
 $1+W_T$ and Case (c) shows $2+W_T$.\\ \\

Fig. 4 Time variation of $(\delta X)^2$, as the atom-field system evolves.
The evolution is shown for three values of detuning, 0.01 (continuous),
0.016061 (dotted) and 0.02 (dashed).  Instants of $(\delta X)^2$ less than
0.5 correspond to squeezing in $X$ quadrature.
\\

Fig. 5  Mandel's $Q$ parameter as a function of time. Instants of $Q < 1$
correspond to sub-Poissonian statistics.  Values of $g$, $k$ and $\bar{n}$ are
same as in Fig. 2 and $\Delta=0.01$. \\ \\

Fig. 6  Overlap of initial and time-evolved field states.  Y-axis corresponds
to $\vert\langle\psi(0)\vert\psi(t)\rangle\vert^2$.  The envelope of the
overlap function decays with time implying that the  initial state is
almost orthogonal to the evolved state.  The values of the parameters
are same as in Fig. 2 and the detuning $\Delta$ is equal to $0.016061$.

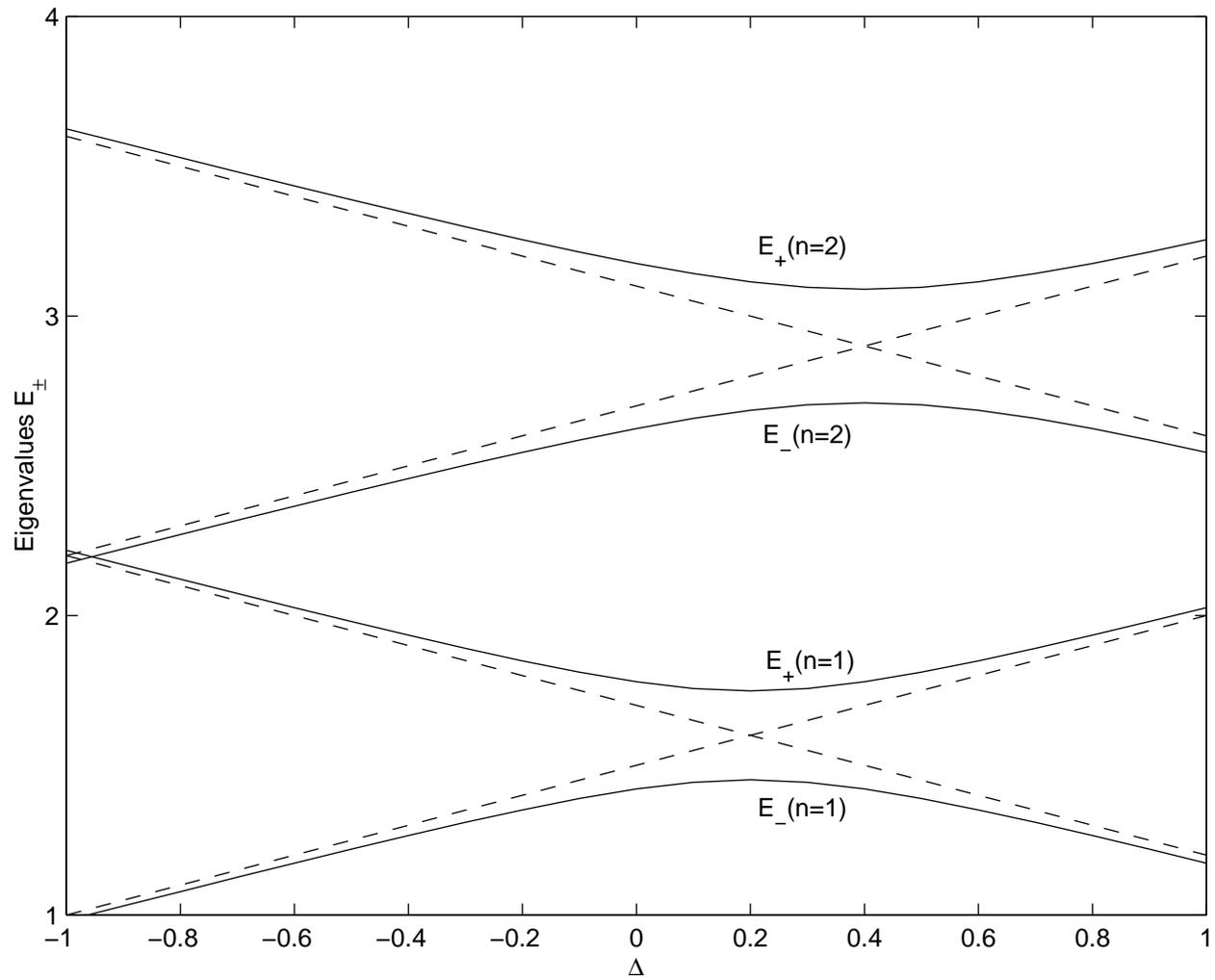
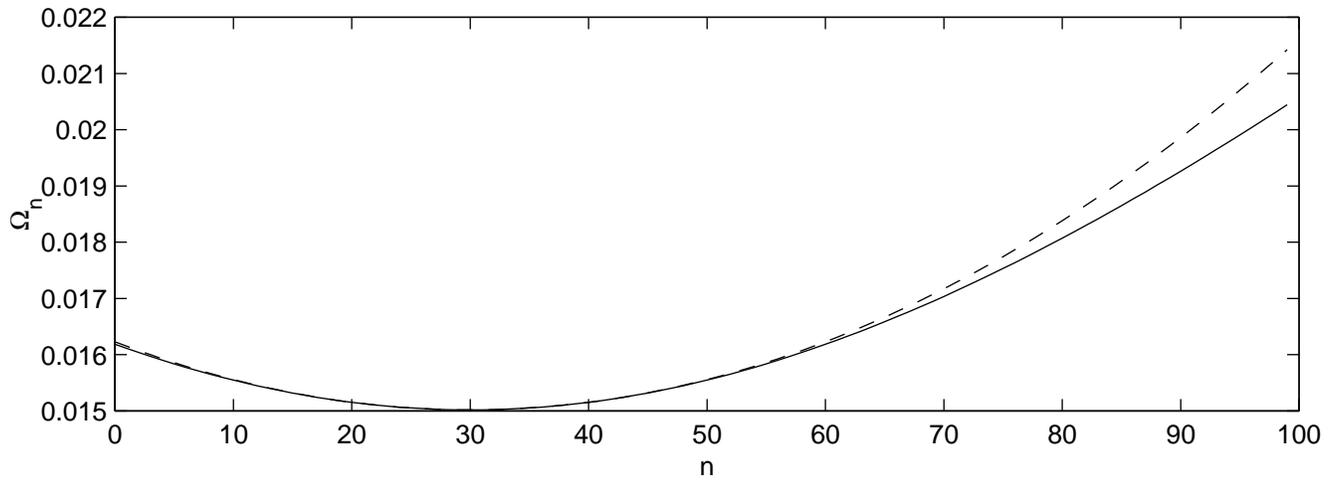
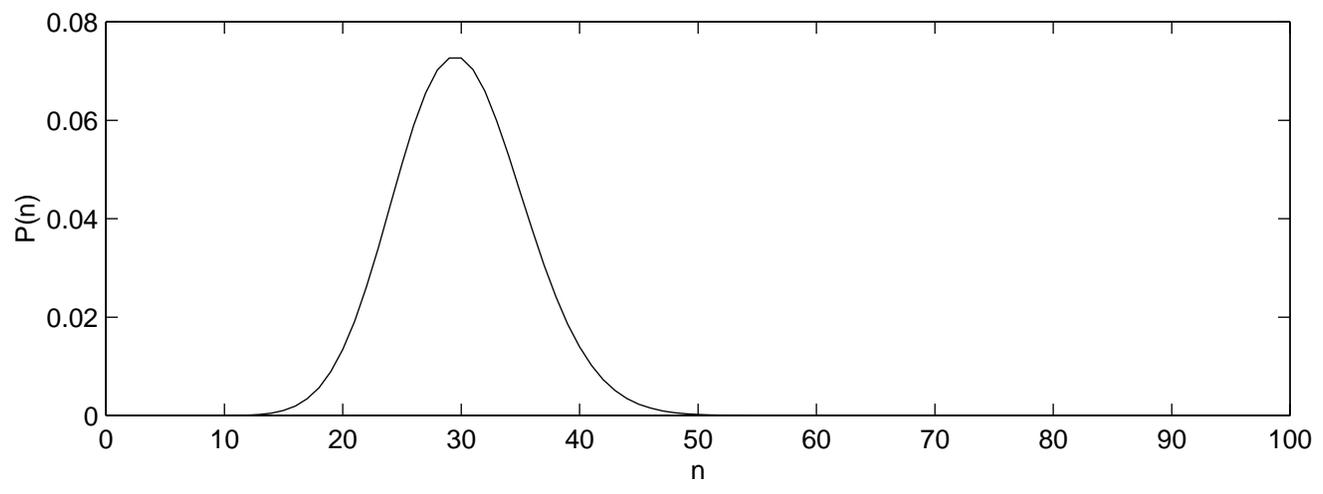
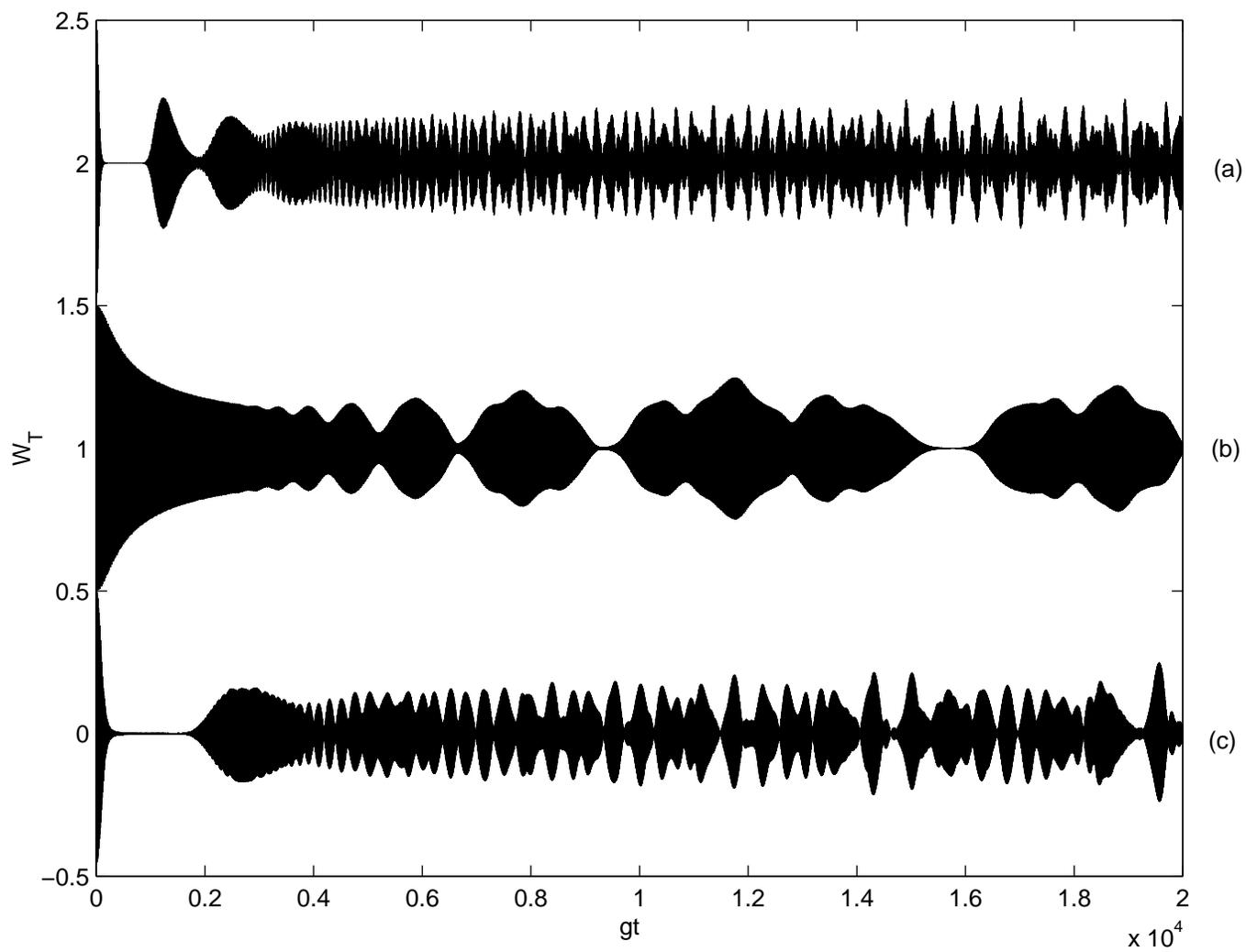
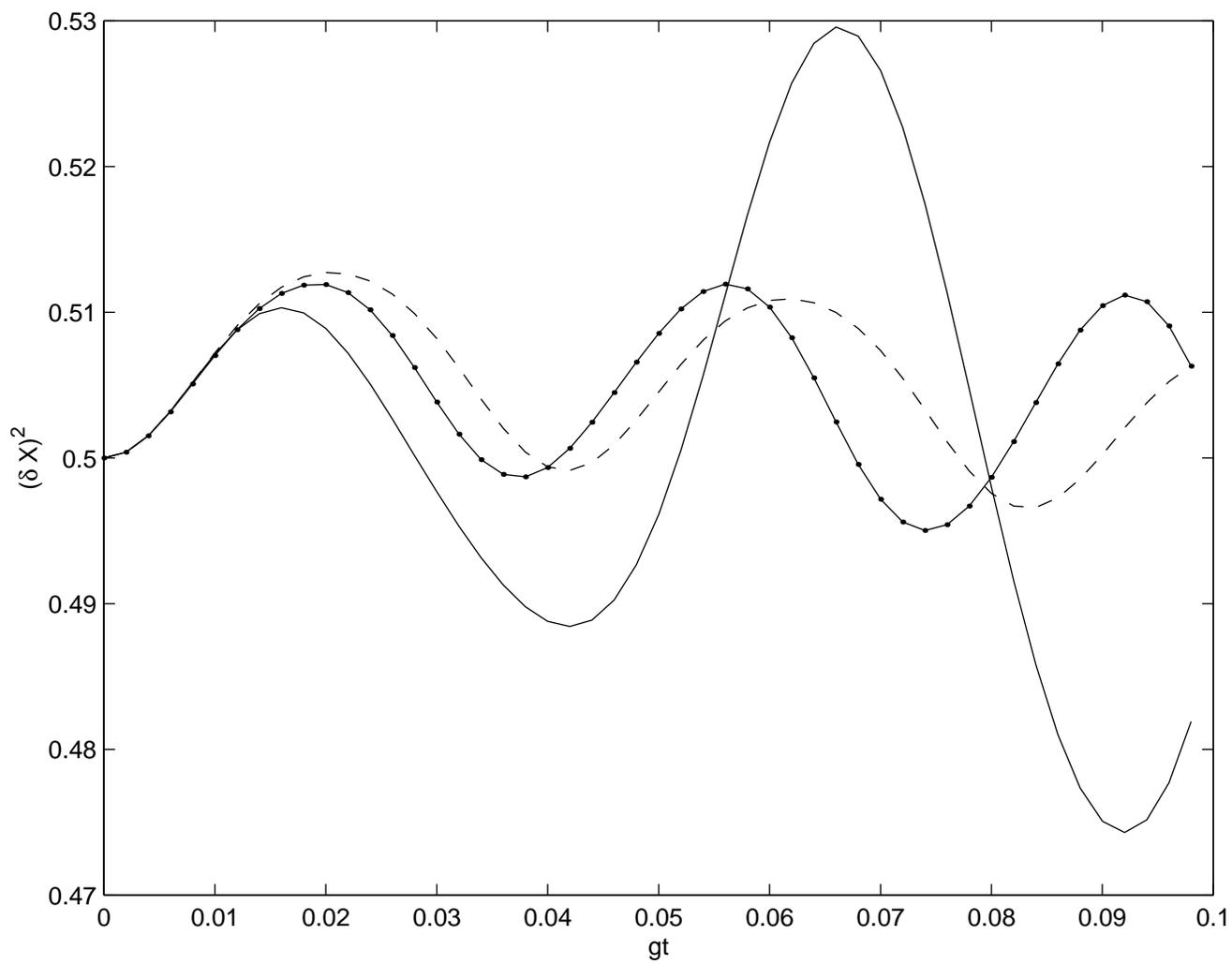
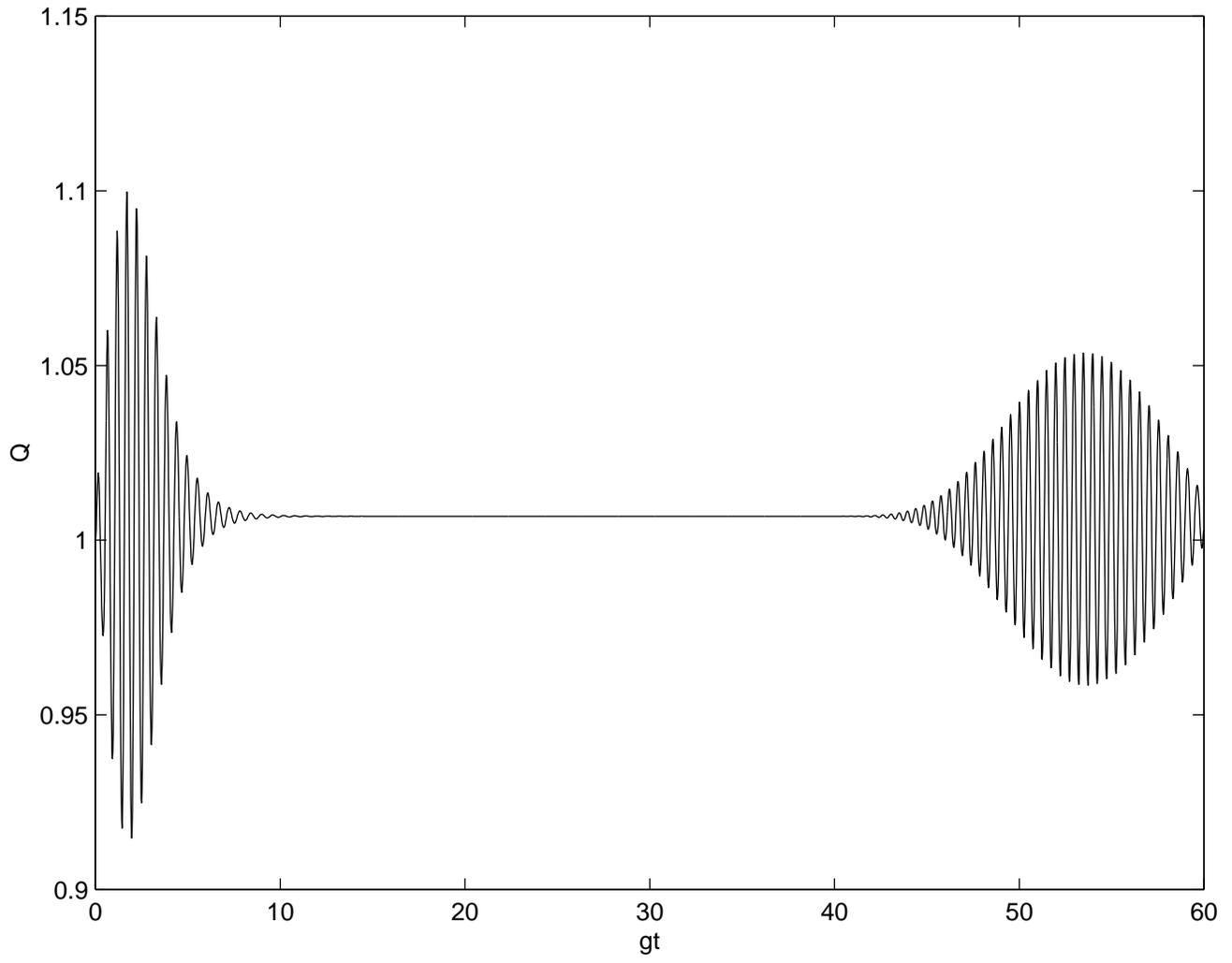
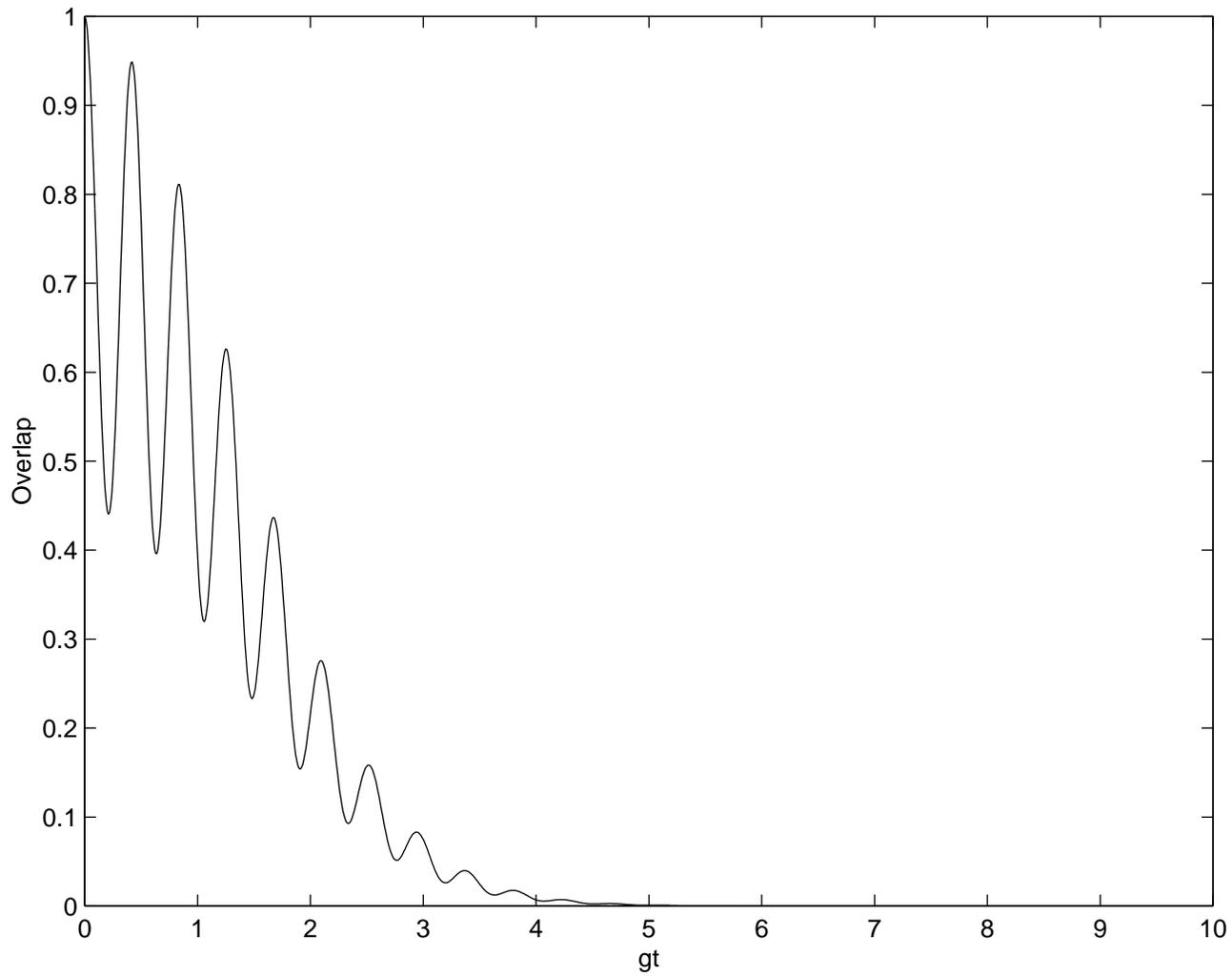
\end{document}